\documentstyle [12pt] {article}
\title{k-MOND (WITH "DARK MATTER" AS A DISTINCTION BETWEEN
 INERTIAL AND GRAVITATIONAL MASS)}

\author{Vladan Pankovi\'c, Darko V. Kapor\\
Department of Physics, Faculty of Sciences, 21000 Novi Sad,\\ Trg
Dositeja Obradovi\'ca 4, Serbia, \\vladan.pankovic@df.uns.ac.rs}

\date {}
\begin {document}
\maketitle \vspace {0.5cm}
\vspace {0.3cm}

\begin {abstract}
In this work we definitely prove a possibility that Milgrom's
modified Newtonian dynamics, MOND, can be consistently interpreted
as a theory with the modified kinetic terms of the usual Newtonain
dynamics, simply called k-MOND. Precisely, we suggest only a
functional dependence between inertial and gravitational mass
tending toward identity in the limit of large accelerations
(characteristic for Newtonian dynamics and its relativistic
generalizations) but which behaves as a principal non-identity in
the limit of small accelerations (smaller than Milgrom's
acceleration constant). This functional dependence implies a
generalization of the kinetic terms (without any change of the
gravitational potential energy terms) in the usual Newtonain
dynamics including generalization of corresponding Lagrange
formalism. Such generalized dynamics, k-MOND, is identical to
Milgrom's MOND. Also, mentioned k-MOND distinction between
inertial and gravitational mass would be formally treated as "dark
matter".
\end {abstract}

In this work we shall definitely prove a possibility that
Milgrom's modified Newtonian dynamics, MOND [1]-[11], can be
consistently interpreted as a theory with the modified kinetic
terms of the usual Newtonain dynamics, simply called k-MOND.
Precisely, we shall suggest only a functional dependence between
inertial and gravitational mass tending toward identity in the
limit of large accelerations (characteristic for Newtonian
dynamics and its relativistic generalizations) but which behaves
as a principal non-identity in the limit of small accelerations
(smaller than Milgrom's acceleration constant). This functional
dependence implies a generalization of the kinetic terms (without
any change of the gravitational potential energy terms) in the
usual Newtonain dynamics including generalization of corresponding
Lagrange formalism. Such generalized dynamics, k-MOND, is
identical to Milgrom's MOND. Also, mentioned k-MOND distinction
between inertial and gravitational mass would be formally treated
as "dark matter". (Name k-MOND is directly inspired by name
k-essence [12]-[14] since in both cases it is supposed that change
of the usual dynamics is realized by change of the usual kinetic
terms by introduction of the more complex kinetic terms. However
any closer correlation between k-essence theories and k-MOND is
not supposed apriorily.)

As it is well-known [1]-[11], for a satisfactory theoretical
explanation of the observationally obtained flat galaxy rotation
curves, Tully-Fisher law etc., Milgrom suggested a modification of
the Newtonian classical dynamics, called MOND.

In MOND it is supposed that there is a new natural constant
representing Milgrom's acceleration ${\it
a}_{M}\simeq10^{-10}\frac {m}{s^{2}}$. It is supposed too that for
system total acceleration vector  {\it {\bf a}} with extremely
small absolute value {\it a}, comparable with Milgrom
acceleration, Newtonian classical dynamical law of the physical
system must be changed in the practically the following form
\begin {equation}
     F \simeq m{\it a} \frac {{\it a}}{{\it a}_{M}}         .
\end {equation}
Here m represents the system mass and {\it a} - the system
acceleration absolute value. For example, for a galaxy with mass
M, that acts by Newtonian gravitational force $F=G\frac
{mM}{R^{2}}$ at a periferically rotating star with mass m, speed v
and small centrifugal acceleration absolute value ${\it a} = \frac
{v^{2}}{R}$ at distance R in respect to galaxy center, where G
represents the Newtonian gravitational constant, MOND dynamics of
this star (1) implies
\begin {equation}
     G\frac {mM}{R^{2}}\simeq m (\frac {v^{2}}{R})^{2} \frac {1}{{\it a}_{M}}
\end {equation}
and further
\begin {equation}
      v^{4} \simeq {\it a}_{M}
\end {equation}
corresponding excellently to Tully-Fisher relation for spiral
galaxies and many other relevant astronomical observational data.

On the other hand it is supposed within MOND that for usual, large
acceleration absolute value, much larger than Milgrom's
acceleration, classical dynamical law must have approximately
usual Newtonian form
\begin {equation}
     F \simeq m {\it a}                       .
\end {equation}

In general case, i.e. for arbitrary small or large acceleration
absolute value, MOND supposes the following dynamical law
\begin {equation}
     F =  m{\it a} \mu(\frac {{\it a}}{{\it a}_{M}})
\end {equation}
where $\mu(\frac {{\it a}}{{\it a}_{M}}$ represents a modification
function depending of {\it a} as the variable and ${\it a}_{M}$ as
the parameter so that for small acceleration absolute value (5)
tends asymptotically toward (1) while for large acceleration
absolute value (5) tends asymptotically toward (4).

Even if MOND simply and elegantly describes important astronomical
observational data physical interpretation or nature of the MOND
is not simple at all. MOND can be an introduction in a quite new
physics or it can be a simple calculation algorithm, i.e.
appropriate approximation formalism. Now we shall suggest the
following physical interpretation of the MOND.

Firstly, we shall suggest the following functional dependence
between inertial mass, mi, and gravitational mass, mg, tending
toward identity in the limit of large accelerations
(characteristic for Newtonian dynamics and its relativistic
generalizations) and behaving as a principal non-identity in the
limit of small accelerations (smaller than Milgrom's acceleration
constant) in the following way
\begin {equation}
     m_{i} =  m_{g} (\frac {{\bf {\it a}}^{2}}{{\bf {\it a}}^{2} + \frac {{\it a}^{2}_{M}}{2^{2}}})^{\frac {1}{2}}=
     m_{g} (\frac {{\it a}^{2}}{{\it a}^{2} + \frac {{\it a}^{2}_{M}}{2^{2}}})^{\frac {1}{2}}=  m_{g} f({\bf {\it a}}^{2})
\end {equation}
where
\begin {equation}
     f({\bf {\it a}}^{2}) = (\frac {{\bf {\it a}}^{2}}{{\bf {\it a}}^{2} + \frac {{\it a}^{2}_{M}}{2^{2}}})^{\frac {1}{2}}
     = (\frac {{\it a}^{2}}{{\it a}^{2} + \frac {{\it a}^{2}_{M}}{2^{2}}})^{\frac {1}{2}}= f({\it a}^{2})
\end {equation}
while ${\it {\bf a}}^{2}={\it a}^{2}$ represents scalar product
between total acceleration vector ${\it {\bf a}}$ with itself. It
can be added that mentioned distinction between inertial and
gravitational mass can be used for the "dark matter" effect
modeling.

Secondly, we shall suppose the following form for system momentum
${\bf p}$ and kinetic energy $E_{k}$
\begin {equation}
     {\bf p} =  m_{i}{\bf v} = m_{g} {\bf v} f({\bf {\it a}}^{2})
\end {equation}
\begin {equation}
     E_{k} = \frac {m_{i}{\bf v}^{2}}{2} =  \frac {m_{g}{\bf v}^{2}}{2} f({\bf {\it a}}^{2})
\end {equation}
where ${\bf v}$ represents the system total velocity so that ${\bf
a} = \frac {d{\bf v}}{dt}$. Obviously these kinematic variables
(that do not include system potential energy and its coordinate
derivations) cab be simply obtained by introduction of (7) instead
usual mass in the corresponding expression in usual Newtonian
dynamics.

Thirdly, suppose that gravitational potential energy of the system
in of the gravitational field of the point like source with mass
$M$ at distance $r$ in respect to system is given by usual form
\begin {equation}
     V_{g} = - \frac {Gm_{g}M}{r}
\end {equation}
so that gravitational force
\begin {equation}
     {\bf F} = -  grad V_{g} = -\frac {Gm_{g}M}{r^{2}}{\bf r_{0}}
\end {equation}
has usual form of a central force, where ${\bf r}= r {\bf r_{0}}$
represents the coordinate vector of the system while ${\bf r_{0}}$
represent unit vector in ${\bf r}$ direction.

Fourthly, suppose that Lagrangian of the system has the following
form
\begin {equation}
   L = E_{k} - V_{g} = \frac {m_{g}{\bf v}^{2}}{2} f({\bf {\it a}}^{2}) + \frac {Gm_{g}M}{r}
\end {equation}
and that dynamical equations of the system can be obtained by
variation calculus of this Lagrangian. These equations, since
Lagrangian depends of ${\bf {\it a}}$, are generalized
Euler-Lagrange, i.e. Euler-Poisson equations
\begin {equation}
  \frac {\partial L}{\partial x_{q}} - \frac {d}{dt} \frac {\partial L}{\partial v_{q}} + \frac {d^{2}}{dt^{2}}\frac {\partial L}{\partial {\it a}_{q}}= 0
   \hspace{0.5cm} {\rm for}  \hspace{0.3cm}  q=1, 2, 3
\end {equation}
or, according to (9), (10), (12),
\begin {equation}
  \frac {\partial V_{g}}{\partial x_{q}} - \frac {d}{dt} \frac {\partial E_{k}}{\partial v_{q}}d/dt + \frac {d^{2}}{dt^{2}}\frac {\partial L}{\partial {\it a}_{q}}= 0
    \hspace{0.5cm} {\rm for}  \hspace{0.3cm}  q=1, 2, 3
\end {equation}
where $x_{q}$, $v_{q}$ and ${\it a}_{q}$ for $q=1, 2, 3$ are
corresponding components of the system radius vector, velocity and
acceleration vector.

Obviously, suggested generalized dynamics of the system (6)-(14)
is in general case different from usual Newton dynamics. This
difference basically originates from (6) only, i.e. only from
difference between inertial and gravitational mass. But
effectively this difference manifests as a change and extension of
the kinetic terms in the usual Newton dynamical equations. For
this reason, as well as for reasons that will be discussed later,
this generalized dynamics will be called k-MOND.

Consider limit of k-MOND in case of large accelerations absolute
value, i.e. for
\begin {equation}
  {\bf {\it a}}^{2}= {\it a}^{2}\gg {\it a}^{2}_{M} > \frac {{\it a}^{2}_{M}}{2^{2}}                   .
\end {equation}
Then, as it is not hard to see, $f({\bf {\i a}}^{2})$ tends toward
1 and suggested generalized dynamics toward usual Newtonian
dynamics in which, as well as in its relativistic generalizations,
there is identity between gravitational and inertial mass, in full
agreement with Eotvos experiments (realized definitely in large
acceleration limit).

Consider, however, opposite limit of the k-MOND in case of small
accelerations, i.e. for
\begin {equation}
  {\bf {\it a}}^{2}= {\it a}^{2}\ll \frac {{\it a}^{2}_{M}}{2^{2}}                           .
\end {equation}

More precisely, consider an especial case of this limit in which
system rotates uniformly. For reason of the formal simplicity,
that does not any diminishing of the general conclusions, we shall
firstly suppose that such uniform rotation exists and then we
shall prove that dynamical equations (13), i.e. (14), are
consistent with such supposition. Also, for reason of similar
formal simplicity we shall suppose that system rotates in $xOy$
plane where $x_{1}=x$ and $x _{2}=y$ while $O$ denotes the
coordinate beginning.

According to mentioned suppositions system radius vector ${\bf r}$
has form
\begin {equation}
  {\bf r} = r(\cos [\omega t] {\bf e_{1}} + \sin [\omega t] {\bf e_{2}} = r {\bf r_{0}}             .
\end {equation}
Here $r$ and $\omega$ represent time independent orbit radius and
angular frequency while ${\bf e_{1}}$ and ${\bf e_{2}}$ represent
unit vectors in $Ox_{1}$ and $Ox_{2}$ directions while
\begin {equation}
  {\bf r_{0}} = \cos [\omega t] {\bf e_{1}} + \sin [\omega t] {\bf
  e_{2}}
\end {equation}
represents time dependent unit vector in ${\bf r}$ direction.
Also, we can define time dependent perpendicular to ${\bf r_{0}}$
tangential unit vector
\begin {equation}
  {\bf r_{t}} = - \sin [\omega t] {\bf e_{1}} + \cos [\omega t] {\bf
  e_{2}}
\end {equation}
so that
\begin {equation}
   \frac {d}{dt} {\bf r_{0}} = \omega {\bf r_{t}}
\end {equation}
and
\begin {equation}
   \frac {d}{dt}{\bf r_{t}} = -\omega {\bf r_{0}}                  .
\end {equation}

Then system velocity ${\bf v}$ has form
\begin {equation}
   {\bf v} = \frac {d}{dt}{\bf r_{t}} = \omega r {\bf r_{t}} = v {\bf r_{t}}
\end {equation}
where velocity absolute value $v = \omega r$ is time independent.
Of course, in any time moment, this velocity is perpendicular to
radius vector so that ${\bf rv}=0$.

Further system acceleration vector has form
\begin {equation}
   {\bf {\it a}} = \frac {d}{dt}{\bf v}= -\omega^{2}r {\bf r_{0}} = -\frac {v^{2}}{r} {\bf r_{0}}  = -{\it a} {\bf r_{0}}
\end {equation}
with time independent absolute value ${\it a}= \frac {v^{2}}{r}$.
Of course, in any time moment, this acceleration vector, parallel
to radius vector, is perpendicular to velocity so that ${\bf {\it
a}v=0}$. Moreover, in any time moment, acceleration vector is
perpendicular to its first time derivation so that ${\bf {it
a}}{\frac {d{\bf {it a}}}{dt}} =0$.

All this implies that system kinetic energy (9) in small
acceleration limit can be approximated by the following expression
\begin {equation}
     E_{k} = m_{g}{\bf v}^{2}  \frac {({\bf {\it a}}^{2})^{\frac {1}{2}}}{{\it a}_{M}} = m_{g}\frac {{\bf v}^{2}}{{\it a}_{M}} (-{\bf {\it a}}{\bf r_{0}})      .
\end {equation}

Then dynamical equations (14) turn out approximately in
\begin {equation}
  \frac {\partial V_{g}}{\partial {\bf r}} - \frac {d}{dt} \frac {\partial E_{k}}{\partial {\bf v}} + \frac {d^{2}}{dt^{2}} \frac {\partial E_{k}}{\partial {\bf {\it a}}} = 0
\end {equation}
or, as it is not hard to see, according to (17)-(23), in
\begin {equation}
  (- \frac {Gm_{g}M}{r^{2}} {\bf r_{0}} ) - (-2\frac {m_{g}}{{\it a}_{M}}\frac {v^{4}}{r^{2}} {\bf r_{0}}) + (\frac {m_{g}}{{\it a}_{M}}{v^{4}}{r^{2}}{\bf r_{0}}) = 0
\end {equation}
and, finally, in
\begin {equation}
  [-\frac {Gm_{g}M}{r^{2}}- (-(\frac {m_{g}}{{\it a}_{M}}{v^{4}}{r^{2}})] {\bf r_{0}} = 0                .
\end {equation}
It implies scalar equation
\begin {equation}
   \frac {Gm_{g}M}{r^{2}}  =   \frac {m_{g}}{{\it a}_{M}}{v^{4}}{r^{2}}
\end {equation}
identical to basic MOND proposition (2).

In this way it is definitely proved a possibility that Milgrom's
modified Newtonian dynamics, MOND, can be consistently interpreted
as a theory with the modified kinetic terms of the usual Newtonain
dynamics (13), (14), simply called k-MOND.

In conclusion we can shortly repeat and point out the following.
In this work we definitely prove a possibility that Milgrom's
modified Newtonian dynamics, MOND, can be consistently interpreted
as a theory with the modified kinetic terms of the usual Newtonain
dynamics, simply called k-MOND. Precisely, we suggest only a
functional dependence between inertial and gravitational mass
tending toward identity in the limit of large accelerations
(characteristic for Newtonian dynamics and its relativistic
generalizations) but which behaves as a principal non-identity in
the limit of small accelerations (smaller than Milgrom's
acceleration constant). This functional dependence implies a
generalization of the kinetic terms (without any change of the
gravitational potential energy terms) in the usual Newtonain
dynamics including generalization of corresponding Lagrange
formalism. Such generalized dynamics, k-MOND, is identical to
Milgrom's MOND. Also, mentioned k-MOND distinction between
inertial and gravitational mass would be formally treated as "dark
matter".

\vspace{1cm}

{\large \bf References}

\begin{itemize}

\item [[1]]  M. Milgrom, The Astrophysical Journal {\bf 270} (1983) 365, 371, 384
\item [[2]] M. Milgrom, {\it Dynamics with A Non-standard Inertia-Acceleration Relation: An  Alternative to Dark Energy in Galactic Systems}, astro-ph/9303012
\item [[3]] M. Milgrom, {\it The Modified Dynamics - A Status Review, General Relativity}, astro-ph/9810302
\item [[4]] M. Milgrom, {\it MOND - A Pedagogical Review}, astro-ph/0112069
\item [[5]] M. Milgrom, {\it MOND - Theoretical aspects}, astro-ph/0207231
\item [[6]] J. D. Bekenstein, {\it Relativistic Gravitation Theory for the MOND Paradigm}, astro-ph/0403694
\item [[7]] J. D. Bekenstein, R. P. Sanders, {\it A Primer for Relativistic MOND theory}, astro-ph/0509519
\item [[8]] J. D. Bekenstein, {\it The Modified Newtonian Dynamics - MOND and Its Implications for New Physics}, astro-ph/0701848
\item [[9]] R. P. Sanders, {\it MOND and Cosmology}, astro-ph/0509532
\item [[10]] R. P. Sanders, {\it From Dark Matter to MOND}, astro-ph/0806.2585
\item [[11]] R. Scarpa, {\it Modified Newtionian Dynamics, An Introductory Review}, astro-ph/0601478
\item [[12]] C. Armanderiz-Picon, T. Damour, V. K. Mukhanov, Phys. Lett. {\bf B 458} (1999) 209, astro-ph/0806.2585
 \item [[13]]  C. Armanderiz-Picon, V. K. Mukhanov, P. Steinhardt, Phys. Rev. Lett. {\bf 85} (2000) 4438, astro-ph/0601478
\item [[14]]  R. de Putter, E. V. Linder, {\it Kinetic k-esence and quintessence}, astro-ph/0705.0400

\end {itemize}

\end {document}